\documentclass[aps,prb,twocolumn,groupedaddress,showpacs]{revtex4}

\usepackage{graphicx}
\usepackage{amsmath}

\begin{document}

\title{Energy gaps in high-$T_c$ superconductors: BCS after all?}

\author{J.L. Tallon$^{1,2}$ and J.G. Storey$^{1}$}

\affiliation{$^1$MacDiarmid Institute, Industrial Research Ltd.,
P.O. Box 31310, Lower Hutt, New Zealand.}

\affiliation{$^2$Quantum Matter Group, Cavendish Laboratory, Cambridge University, CB3 0HE, United Kingdom.}

\date{\today}

\begin{abstract}

\end{abstract}


\maketitle

{\bf A major impediment to solving the problem of high-$T_c$ superconductivity is the ongoing confusion about the magnitude, structure and doping dependence of the superconducting gap, $\Delta_0$, and of the mysterious pseudogap found in underdoped samples\cite{TallonLoram}. The pseudogap opens around the ($\pi$,0) antinodes below a temperature $T^*$ leaving Fermi arcs across the remnant Fermi surface\cite{Kanigel} on which the superconducting gap forms at $T_c$. One thing that seems agreed is that the ratio $2\Delta_0/k_BT_c$ well exceeds the BCS value and grows with underdoping\cite{Miyakawa1,Miyakawa2}, suggesting unconventional, non-BCS superconductivity. Here we re-examine data from many spectroscopies, especially Raman $B_{1g}$ and $B_{2g}$ scattering\cite{Sacuto,Guyard}, and reconcile them all within a two-gap scenario showing that the points of disagreement are an artefact of spectral-weight loss arising from the pseudogap. Crucially, we find that $\Delta_0(p)$, or more generally the order parameter, now scales with the mean-field $T_c$ value, adopting the weak-coupling BCS ratio across the entire phase diagram.}

\begin{figure*}
\centering
\includegraphics*[width=90mm, angle=270]{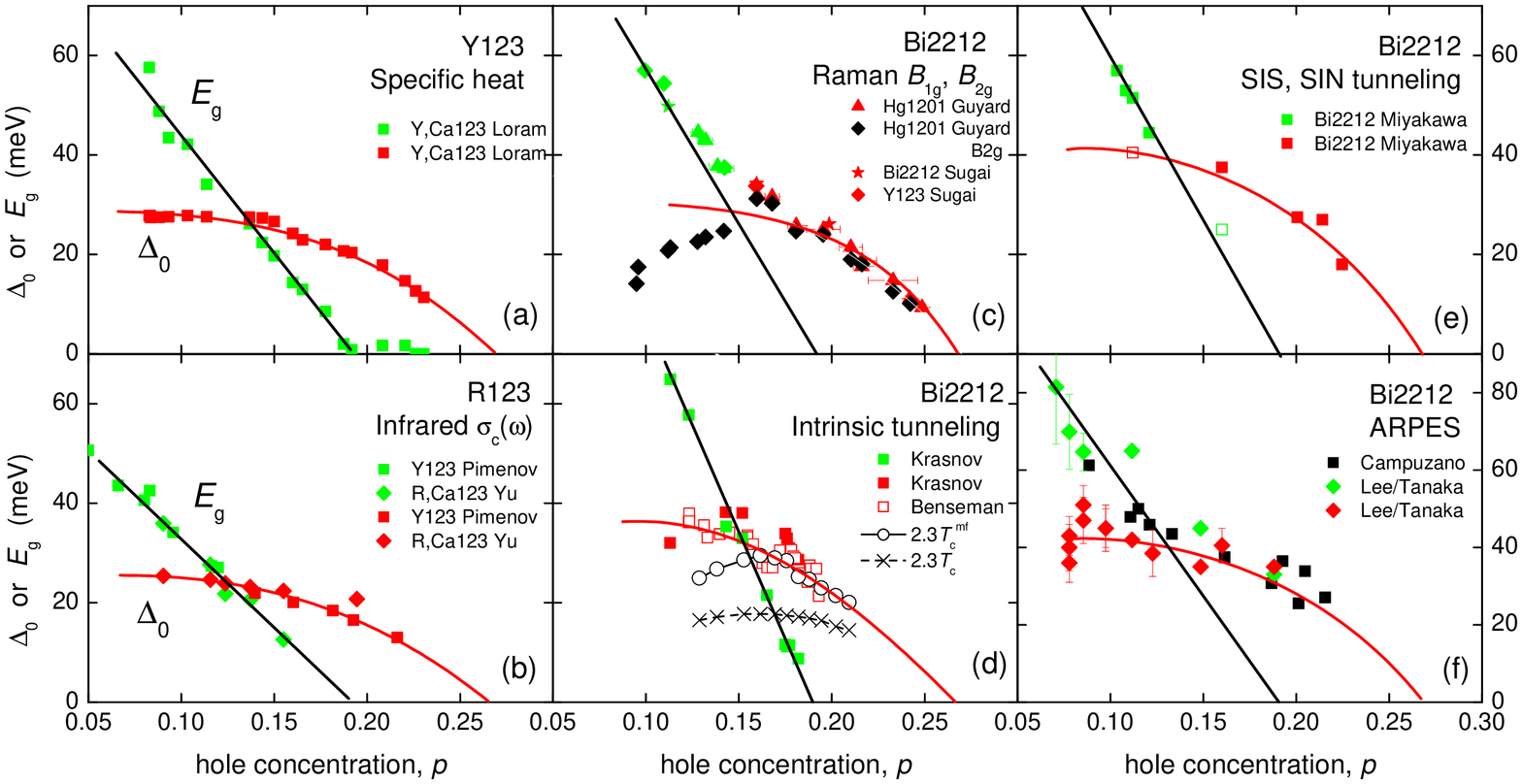}
\caption{\small The doping dependence of the maximum $d$-wave SC gap, $\Delta_0$, and the pseudogap, $E_g$. Green data
points show the gap observed above $T_c$ (i.e. the pseudogap) while
red data points show the gap observed below $T_c$ (with care, the SC
gap). (a) from specific heat measurements for
Y$_{0.8}$Ca$_{0.2}$Ba$_2$Cu$_3$O$_{7-\delta}$; (b) from infrared
$c$-axis conductivity for
R$_{0.8}$Ca$_{0.2}$Ba$_2$Cu$_3$O$_{7-\delta}$; (c) from Raman
$B_{1g}$ and $B_{2g}$ measurements on three 90K superconductors; (d) from intrinsic tunneling, (e) from SIS/SIN
tunneling, and (f) from ARPES. The open circles in panel (d) are values of 2.5$k_BT_c^{mf}$.}
\label{SHgaps}
\end{figure*}

We have long argued\cite{TallonLoram} from thermodynamic and Knight-shift data that there are two distinct energy gaps in the electronic density of states (DOS). These are plotted in Fig.~\ref{SHgaps}(a) from the specific heat data for Y$_{0.8}$Ca$_{0.2}$Ba$_2$Cu$_3$O$_{7-\delta}$\cite{Loram}. (Similar results were found for Bi$_{2}$Sr$_{2}$CaCu$_2$O$_{8+\delta}$\cite{Loram} (Bi-2212)). $\Delta_0$ is the projected maximum in the $T$=0, $d$-wave SC gap, $\Delta({\bf k})$, at the ($\pi$,0) antinode. $\Delta_0(p)$ falls monotonically with increasing doping, and vanishes at about $p=0.27$ holes/Cu at the superconductor/metal transition. $E_g$ is the magnitude of the pseudogap (PG) at the antinode, where, at the time, we assumed a triangular gap in the DOS which fills with increasing temperature. This filling effectively reflects the growth of the Fermi arcs with increasing temperature\cite{Kanigel,StoreyFArc}. $E_g$ also decreases with doping but, unlike $\Delta_0$, falls abruptly to zero at $p_{crit}=0.19$. Subsequent studies\cite{Fluc} show that the $E_g(p)$ doping dependence is probably sublinear with $E_g(p)\propto (p_{crit}-p)^{0.8}$.

It is important next to consider the \textbf{k}-dependence of these gaps. The SC gap has the $d$-wave form, $\Delta(\theta)=\Delta_0\cos(2\theta)$ with nodes at $\theta$=45$^{\circ}$ where $\theta$ is the angle around the Fermi surface (FS). In contrast, as shown by specific heat\cite{StoreyFArc}, Raman\cite{StoreyFArc}, NQR\cite{Zheng}, NMR\cite{Mali}, STM\cite{Pushp} and ARPES\cite{StoreyFArc}, the ground-state (GS) PG is not nodal but closes at $\theta_0$ ($<45^{\circ}$) leaving remnant Fermi arcs around the nodes. Thus\cite{StoreyFArc}
\begin{equation}
E_g(\theta)=
\begin{array}{ll}
E_g^0\cos{\left(\frac{2\pi\theta}{4\theta_{0}}\right)} & (\theta<\theta_{0})\\
\end{array},
\label{PGEQ}
\end{equation}
where $\theta_0$ defines the end of the Fermi arc. It is on the remnant Fermi arc that the SC gap opens. With decreasing doping the GS Fermi arc narrows and only closes at the node as $p\rightarrow0.05$ where superconductivity disappears. Here there is no remnant arc on which the SC gap may appear and the loss of superconductivity at that point follows naturally.

\begin{figure}
\centerline{\includegraphics*[width=85mm]{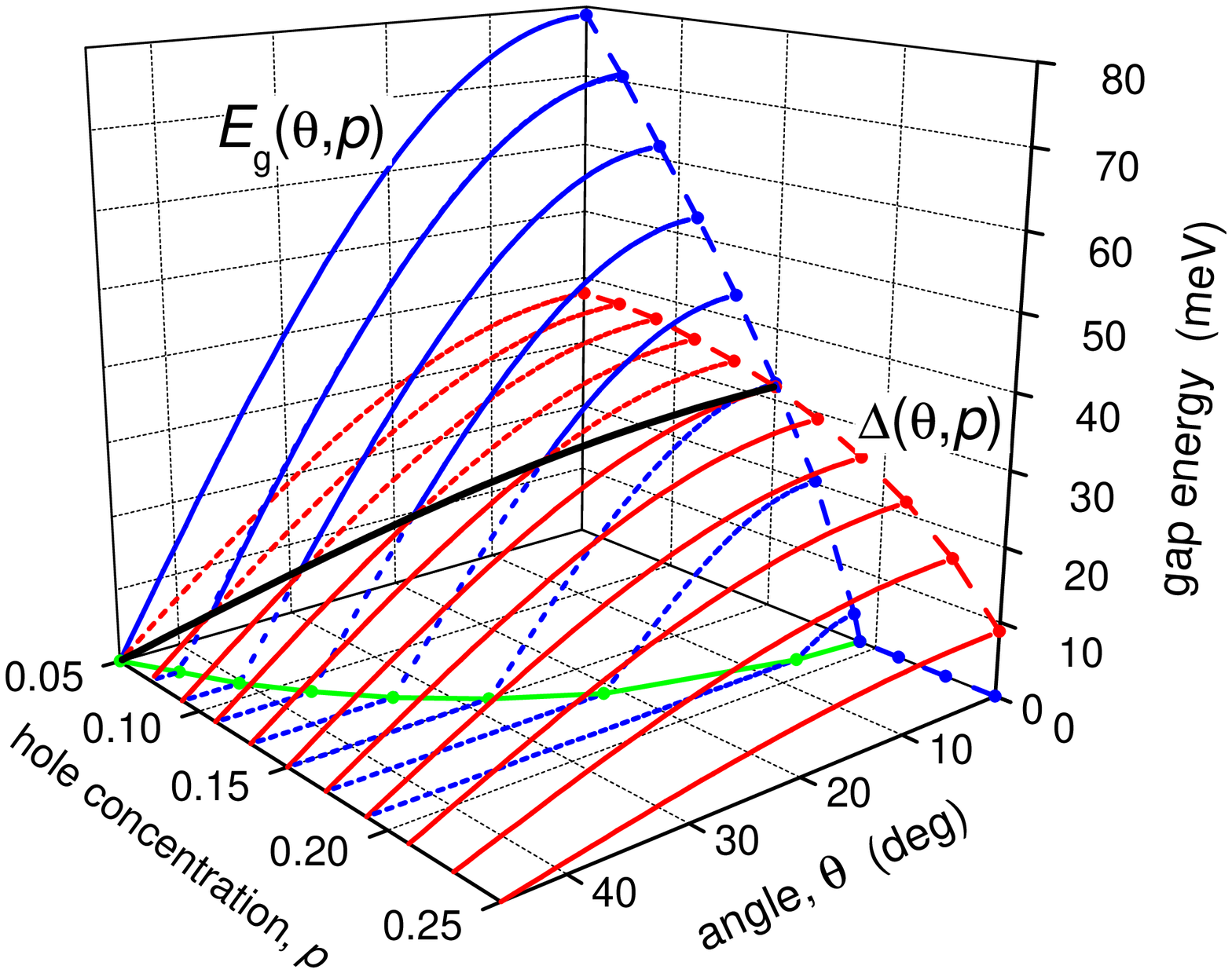}}
\caption{\small Schematic diagram, based on Fig.~\ref{SHgaps} showing the doping- and angular-dependence of the combined gap structure for the SC gap, $\Delta_0(\theta)$ (red curves), and the pseudogap, $E_g(\theta)$ (blue curves), where $\theta$ is the angle around the Fermi surface. The dotted blue curves show the ground-state pseudogap behavior in the absence of superconductivity. The green curve on the zero-energy plane is the contour of doping dependent ground-state Fermi arcs. The black curve shows the contour where $E_g(\theta,p)=\Delta(\theta,p)$.}
\label{gapstructure}
\end{figure}

We plot in Fig.~\ref{gapstructure} the doping and angular dependence of $\Delta$ and $E_g$. The GS magnitudes are averaged from all the data shown in Fig.~\ref{SHgaps}. The figure also shows (green curve on the zero-energy plane) the evolution of $\theta_0$ with doping as deduced previously\cite{StoreyFArc}, ranging from $\theta_0=0$ at $p=0.19$ where the PG first opens, to $\theta_0=45^{\circ}$ at $p=0.05$. The black curve is the contour of equality of the two gaps. [Note that at $p=0.15$ the two gaps are equal at the antinode but the PG has removed all the states below the dashed blue PG curve that would otherwise be pushed up above the SC gap. Thus there is a loss of spectral weight that is reflected e.g. in a marked diminution of the $T=0$ superfluid density\cite{Bernhard}. This persists all the way out to $p=0.19$, where the PG finally closes\cite{Bernhard}. Importantly, this observed loss of GS superfluid density illustrates that the PG is unrelated to thermal phase fluctuations because at $T=0$ all thermal fluctuations vanish.]

The impact of this gap structure on the DOS is illustrated by the calculated DOS for Tl$_2$Ba$_2$CuO$_{6+\delta}$ (see below) shown in Fig.~\ref{DOS}. Three cases are shown: with the PG only, the SC gap only, and both coexisting. The loss of spectral weight in the coherence peak due to the PG is evident, as is the shift in the SC gap feature from $\Delta_0$ with no PG to $\sqrt{\Delta_0^2+E_g^2}$ with the PG. The calculated DOS (red curve) corresponds in detail to the tunneling DOS observed in recent STM data\cite{Pushp}. The van Hove singularity (vHs) can be seen at $E - E_F\approx$ -85 meV.

This is our starting point. The question we address here is: are other techniques consistent with this phenomenology? As a general principle the PG is best observed above $T_c$ while the SC gap, in the absence of fluctuations, is observed only below $T_c$. If, and when, the PG and SC gap coexist below $T_c$ care is required to distinguish them. As a visual aid, in all data shown in Fig.~\ref{SHgaps} gaps measured above $T_c$ (the PG) are shown in green, while gaps measured only below $T_c$ are shown in red.

We turn then to our first comparison, that of the $c$-axis infrared conductivity. We show in Fig.~\ref{SHgaps}(b) the two distinct gaps observed in ellipsometry studies\cite{Yu,Pimenov}. The green data points show the PG determined from the loss of spectral weight below a
frequency 2$E_g/\hbar$ which begins already around 300K, while the
red data points show the SC gap determined from the loss of spectral
weight below a frequency 2$\Delta_0/\hbar$ which begins only below
$T_c$. In the heavily overdoped region only the SC gap is observed. Most
notably there is a range of doping in which both gaps are clearly seen to
coexist, and where the relative magnitudes of $E_g(p)$ and
$\Delta_0(p)$ swap over. Both gaps closely match the specific-heat-derived gaps in panel (a). This is very encouraging, but when we turn to Raman data the result is rather different, at least superficially.

The recent revival of the two-gap scenario for cuprate superconductivity is due largely to the insights derived from Raman scattering in the B$_{1g}$ and B$_{2g}$ modes\cite{Sacuto,Guyard}. But while the B$_{1g}$ gap was found to decrease monotonically from a large magnitude comparable with the exchange interaction, $J$, the B$_{2g}$ gap more or less followed the dome-shaped $T_c(p)$ phase curve. Typical data is shown in Fig.~\ref{SHgaps}(c) for Hg-1201\cite{Guyard}, Bi-2212\cite{Sugai} and Y-123\cite{Sugai}, each having similar maximum $T_c$ values. Again, the $B_{1g}$ gaps are shown in green and red depending on whether they are observed above or only below $T_c$, respectively. The $B_{2g}$ gap is shown in black diamonds. Like the infrared gaps these are pairbreaking gaps so we have divided by a factor 2. The following features are apparent: (i) the $B_{1g}$ gap, in its $p$-dependence, is like $E_g$ in panels (a) and (b) but is actually larger; and (ii) as noted, the $B_{2g}$ gap is not monotonic but follows the SC dome seen in $T_c(p)$. These features led the Sacuto group to interpret the underdoped $B_{1g}$ gap as the PG and the $B_{2g}$ gap as a direct measure of the SC gap\cite{Sacuto,Guyard}. H\"{u}fner {\it et al}.\cite{Hufner}, in an influential review, draw the same conclusion. Thus, at face value, the $B_{1g}$ and $B_{2g}$ gaps appear to contradict the infrared and specific heat data. However, the analysis assumes a single-gap model to interpret spectra in a two-gap scenario. Below, we reconcile all these data.

As shown in the inserts to Fig.~\ref{RamanPG}(a) and (b), the B$_{1g}$ scattering symmetry probes around the gap antinodes at ($\pi$,0), while B$_{2g}$ scattering is dominated by contributions near the $d$-wave gap nodes. Where there is a single $d$-wave SC gap the Raman scattering peak in $B_{1g}$ is a pairbreaking peak at 2$\Delta_0$ and at $\sim$$\sqrt{2}\Delta_0$ for $B_{2g}$. The $\sqrt{2}$ factor is for a circular FS. For a typical cuprate FS it is $\approx 4/3$. On the other hand, where there is a second coexisting gap centred on the antinodes, as in the case of a PG with finite Fermi arcs, this removes the pile-up of states above the SC gap near ($\pi$,0) that contributes to the 2$\Delta_0$ pair-breaking peak, pushing the $B_{1g}$ gap to higher energy and the $B_{2g}$ gap to lower energy\cite{StoreyRaman}. This merely reflects the change in DOS shown in Fig.~\ref{DOS}. The scattering intensity also falls. In short, the $B_{2g}$ pair-breaking gap is not the SC gap, $\Delta_0$ and it is somewhat fortuitous that it roughly follows the $T_c(p)$ phase curve.

\begin{figure}
\centerline{\includegraphics*[width=75mm]{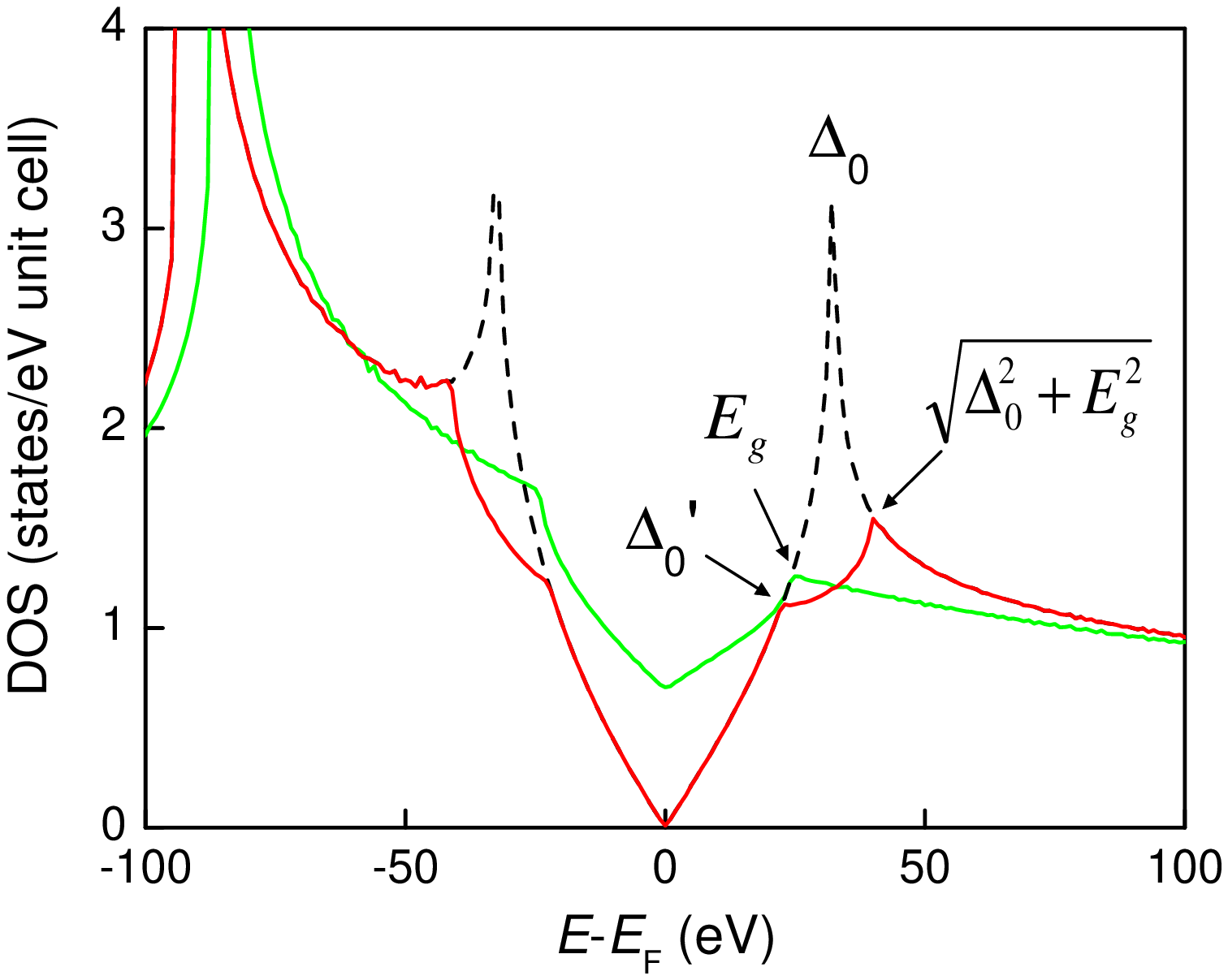}}
\caption{\small The calculated DOS based on a rigid ARPES-derived dispersion for Tl$_2$Ba$_2$CuO$_6$ for three cases: with PG only (green), with SC gap only (black), and with both (red) where the DOS acquires two gap features and loses spectral weight in the coherence peak. The sub-gap feature marked $\Delta_0^{\prime}$ causes the $B_{2g}$ gap and approximates the order parameter. The vHs lies at $E - E_F\approx$ -85 meV.}
\label{DOS}
\end{figure}

We have calculated the detailed Raman $B_{1g}$ and $B_{2g}$ response and the results are summarised in Fig.~\ref{RamanPG}. The imaginary part of the unscreened non-resonant Raman response at $T=0$ is given by\cite{RAMAN}
\begin{equation}
\chi^{''}_{0}\left(q=0,\omega\right)=\int\frac{d^2k}{\left(2\pi\right)^2}\delta\left(\omega-2E(\textbf{k})\right)\frac{|\Delta(\textbf{k})|^2}{E(\textbf{k})^2}|\gamma(\textbf{k})|^2
\label{RAMANEQ}
\end{equation}
where the integral is over occupied states below $E_F$, $\Delta(\textbf{k})=\frac{1}{2}\Delta_{0}\left(\cos{k_{x}}-\cos{k_{y}}\right)$ is the $d$-wave SC gap function and
$E(\textbf{k})=\sqrt{\epsilon(\textbf{k})^2+|\Delta(\textbf{k})|^2}$. In the $B_{1g}$ scattering symmetry
$\gamma(\textbf{k})^{B_{1g}}=\gamma{}B_{1g}\left(\cos{k_{x}}-\cos{k_{y}}\right)$. This function is plotted in the inset to Fig.~\ref{RamanPG}(a) and, as noted, is maximal on the antinodal sections of the FS. For $B_{2g}$, $\gamma(\textbf{k})^{B_{2g}}=\gamma{}B_{2g}\sin{k_{x}}\sin{k_{y}}$ and the response, shown in the inset to Fig.~\ref{RamanPG}(b), is mainly nodal. To compare with the HgBa$_2$CuO$_5$ data in Fig.~\ref{SHgaps}(c), we use a rigid dispersion reported for Tl$_2$Ba$_2$CuO$_6$ from fitted ARPES data\cite{Plate}. The PG is modeled with doping-dependent Fermi arcs using Eq.~\ref{PGEQ}\cite{StoreyFArc}. We fit the parameter values for $E_g(p)$ and $\theta_{0}$ by assuming $\Delta_0$ and reproducing the $B_{1g}$ and $B_{2g}$ gaps, respectively, shown in Fig.~\ref{SHgaps}(c). We also calculate spectra in the absence of the PG. The hole concentration is obtained from integrating the DOS.

\begin{figure*}
\centering
\includegraphics[width=115mm]{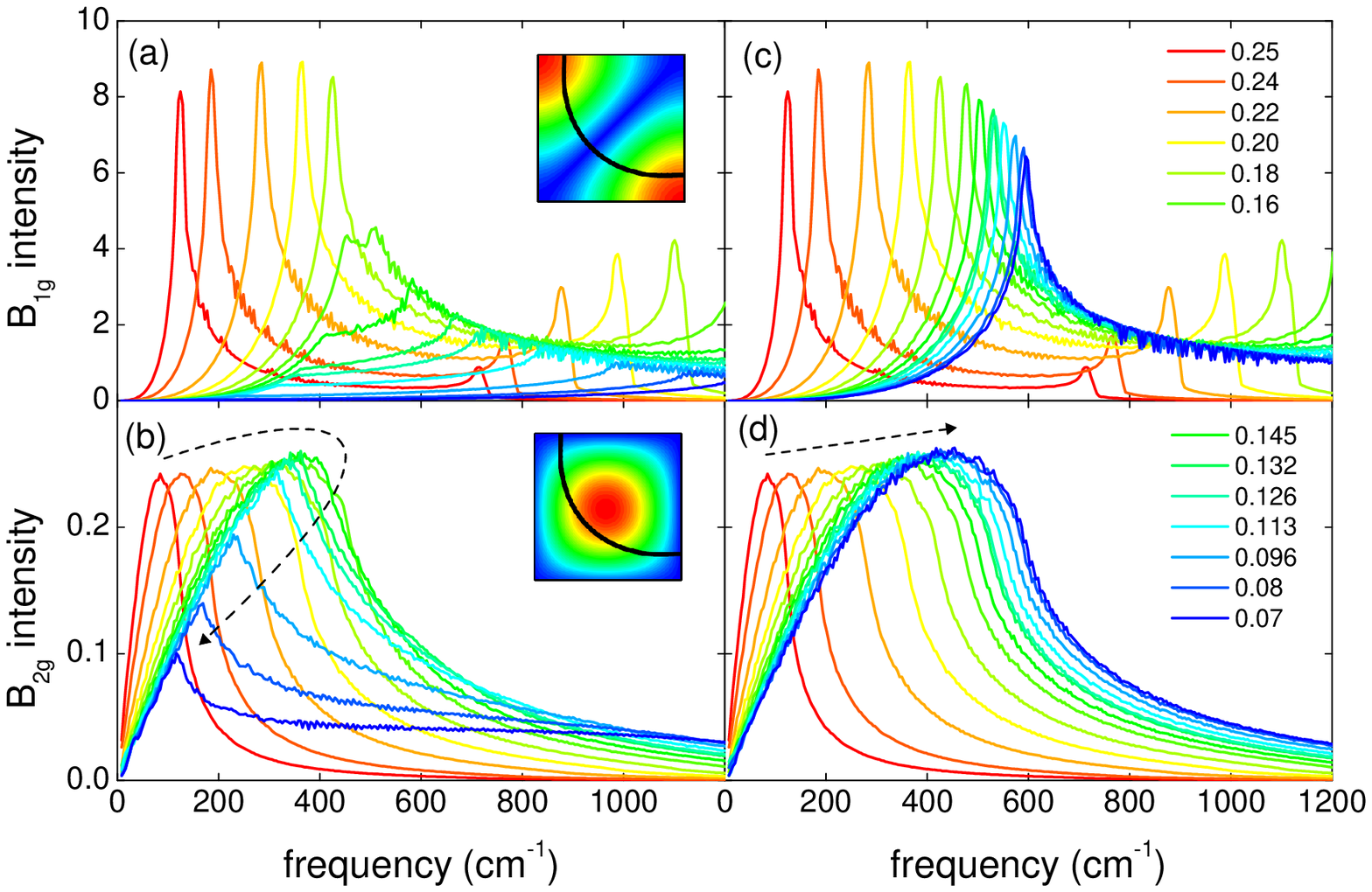}%
\hspace{1mm}%
\includegraphics[width=63mm]{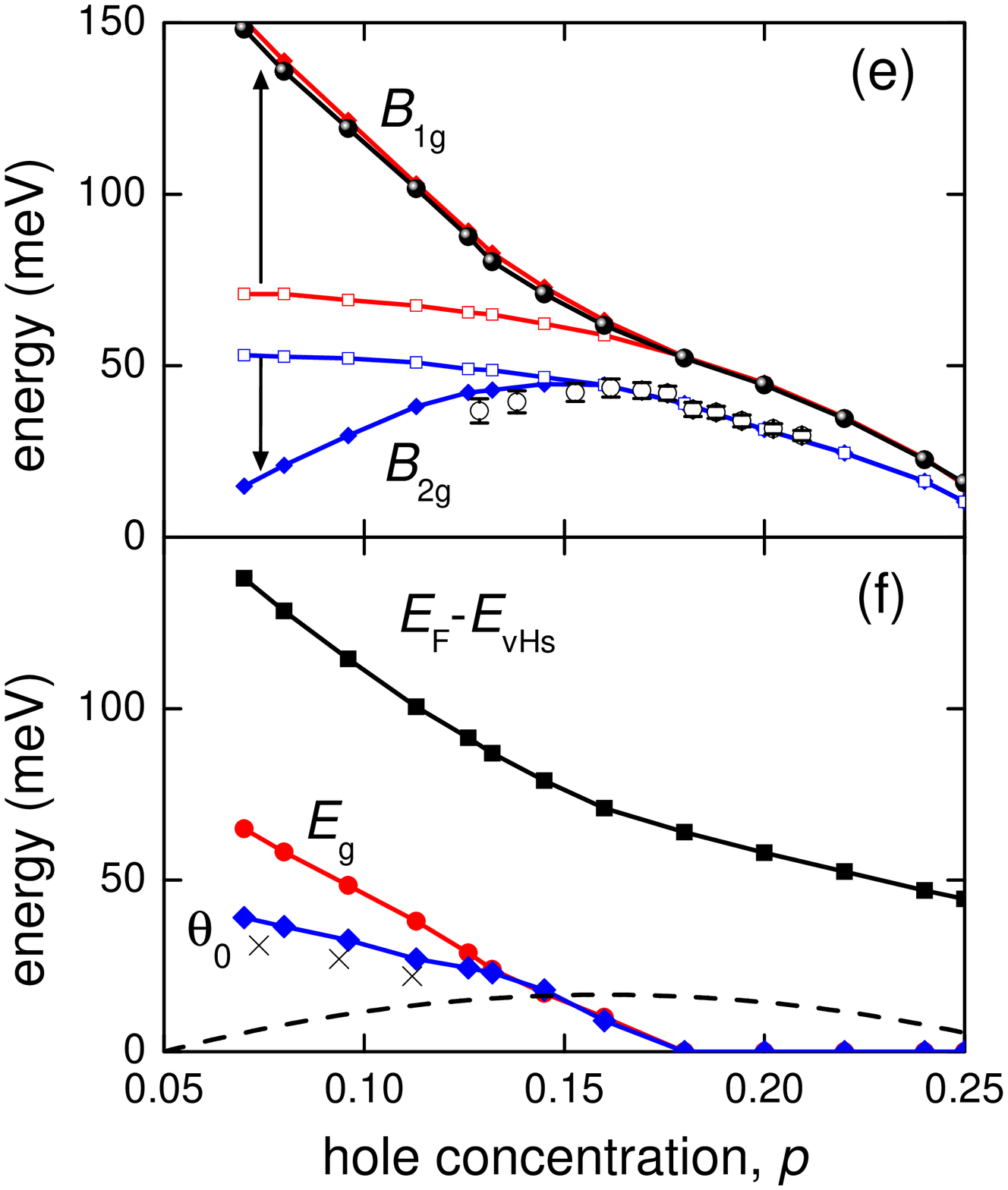}
\caption{\small
The calculated Raman response for HgBa$_2$CuO$_4$ for various doping levels listed in the legends of panels (c) and (d): (a) and (b) with the PG, (c) and (d) without the PG. Insets show $\gamma(\textbf{k})$ from Eq.~\ref{RAMANEQ} plotted in the first quadrant of the Brillouin zone for $B_{1g}$ (a) and for $B_{2g}$ (b). Solid black curves are the Fermi surface.  Dashed arrows in (b) and (d) contrast the rise and fall in $B_{2g}$ gap in the presence of a PG with the monotonic behaviour without a PG. Panel (e) shows $B_{1g}$ (red) and $B_{2g}$ (blue) pair-breaking gap values with (solid symbols) and without (open symbols) the PG. Arrows indicate the effect of the PG. Black symbols are 2$\sqrt{\Delta_0^2 + E_g^2}$. Open black circles with error bars are values of 4.3$k_BT_c^{mf}$ divided by 4/3, the renormalisation factor between the $B_{1g}$ and $B_{2g}$ gaps. (f) Parameter values obtained from the data fits. Crosses show $\theta_0$ determined from STM\cite{Pushp}.} \label{RamanPG}
\end{figure*}

Fig.~\ref{RamanPG} shows the calculated $B_{1g}$ and $B_{2g}$ spectra with (panels (a) and (b)) and without ((c) and (d)) the PG. Leaving aside the Raman continuum all features correspond in detail to those reported in Fig.~\ref{SHgaps}(c) and in the raw spectra\cite{Sacuto}, including the reduction in peak intensity for $B_{2g}$. In the absence of the PG both modes progressively shift to higher energy with decreasing doping whereas, with the PG, the peak in the $B_{2g}$ mode rises then falls (and weakens) as shown by the dashed arrow. This reflects the impact of the PG in removing antinodal spectral weight. At the same time the $B_{1g}$ gap continues to rise, though more rapidly than in the absence of the PG. The peak energies are plotted in Fig~\ref{RamanPG}(e) and the fitting parameters $E_g$ and $\theta_0$ are plotted in Fig~\ref{RamanPG}(f).

We note the following trends. In the absence of the PG the $B_{1g}$ and $B_{2g}$ gap features track each other with a ratio of 4/3, as expected, and they precisely track $\Delta_0$ as determined from the peak-to-peak gap in the DOS. With the introduction of the PG the $B_{1g}$ antinodal gap feature continues to take the magnitude of the SC gap in the heavily overdoped region beyond $p_{crit}$, but below $p_{crit}$ it rises rapidly and its magnitude actually exceeds $E_g$ just as we have noted. Here the $B_{1g}$ gap is 2$\sqrt{\Delta_0^2 + E_g^2}$ which is plotted by the black symbols in Fig.~\ref{RamanPG}(e).

The vHs is seen in panels (a) and (c) by the second (high-energy) peak in $B_{1g}$. Experimentally, this would be concealed by the electronic continuum above the gap but could emerge from the electronic background at high doping near or beyond the vHs. A possible observation of the vHs in $B_{1g}$ Raman scattering at high doping is reported by Limonov {\it et al.}\cite{Limonov}.

Turning to the $B_{2g}$ gap it is notable that this trends to zero at the SC/insulator boundary at $p\approx0.05$ and again one can understand how the $B_{2g}$ gap could be confused with $\Delta_0$. But it falls to zero because, as shown, the ground-state Fermi arc shortens with decreasing doping, collapsing to the node at  $p\approx0.05$. The residual DOS pile-up is then pushed all the way to the node, weakening as it approaches, and vanishes there. This result is borne out again in the present data fits - see Fig.~\ref{RamanPG}(f) where $\theta_0\rightarrow45^{\circ}$ as $p\rightarrow0.05$. Note that if the GS of the PG were always nodal then our calculations show that the $B_{2g}$ gap follows the $B_{1g}$ gap (reduced by the renomalisation factor of 3/4). It is the non-monotonic behavior of the $B_{2g}$ gap that confirms the finite Fermi arc in the GS PG. The $B_{2g}$ pair-breaking gap can be approximated by $\frac{3}{2}\sqrt{\Delta_0^2 - (\frac{2}{3}E_g)^2}$ and roughly follows the magnitude of the SC order parameter, $\Delta_0^\prime$. In general $\Delta_0^\prime < \Delta_0$ but in the absence of the PG they are equal.

The deduced values of $\Delta_0$, $E_g$ and $\theta_0$ closely match those derived from specific heat. For $\theta_0$, compare the green curve in Fig.~\ref{gapstructure} with blue diamonds in Fig.~\ref{RamanPG}(f). $\theta_0$ values extracted from recent STM data on Fermi arcs\cite{Pushp} show the same trend (crosses in Fig.~\ref{RamanPG}(f)). We conclude that there is remarkable consistency between the specific heat, NMR Knight shift, c-axis infrared conductivity and Raman $B_{1g}$ and $B_{2g}$ scattering.

We now turn to tunneling gaps - firstly intrinsic tunneling. The
results of intrinsic tunneling would appear to be regrettably
downplayed. Typically involving up to 20 bilayer stacks, intrinsic
tunneling is arguably a bulk tunneling technique in comparison with
scanning tunneling spectroscopy which just probes the outer CuO$_2$ layer. Intrinsic tunneling consistently
reveals very sharp coherence peaks\cite{Krasnov,Yurgens} and the
presence of two gaps with very different doping and temperature
dependences. For example, investigations on
Bi$_2$(Sr$_{2-x}$La$_x$)CuO$_{6+\delta}$ reveal a SC gap that closes
at $T_c$ with $2\Delta_0/k_BT_c = 4.2$ while the pseudogap remains
fixed in value with increasing temperature and fills rather than
closes\cite{Yurgens}. With increasing doping the pseudogap reduces
in magnitude and falls to zero at $p\approx0.20$ holes/Cu. This
phenomenology is wholly consistent with what has been described above. In the case of
Bi-2212 the evolution of $E_g$ and $\Delta_0$ with doping has been reported in
detail by Krasnov and coworkers\cite{Krasnov}. This system shows a similar behavior to Bi-2201 discussed
above and the gap values are plotted in Fig.~\ref{SHgaps}(d) by the solid symbols. Here each gap
can be discerned, both when $E_g > \Delta_0$ and when $E_g <
\Delta_0$. The detailed variation with doping is again
consistent with all the data shown in panels (a) to (c),
with the pseudogap falling abruptly to zero at $p=0.19$. Such data
has been questioned on the basis of overheating of the nanoscale
mesas\cite{Zavaritsky} but this can be addressed and
eliminated\cite{Krasnov2,Krasnov3}. More recent intrinsic tunneling studies by Benseman {\it et al.}\cite{Benseman} have been carried out in closely-spaced increments in doping. Their values of $\Delta_0$ are also plotted in Fig.~\ref{SHgaps}(d) by the open symbols. They fully confirm the Krasnov data.

SIN and SIS break-junction tunneling spectra have been measured for
Bi-2212 by Miyakawa {\it et al.}\cite{Miyakawa1,Miyakawa2} and their low-$T$ gaps are plotted by the solid symbols in Fig.~\ref{SHgaps}(e). These authors drew attention to the very large gaps observed in the underdoped region where 2$\Delta_0/k_BT_c$ was found to progressively grow with underdoping (reaching 18.9 for a $T_c$=70K sample). Because these gaps tended to decrease near $T_c$ they were presumed to be SC gaps which profoundly exceeded BCS weak-coupling behavior. Unlike all other panels in Fig.~\ref{SHgaps} here the green and red symbols do not represent gaps reported above and below $T_c$ respectively. No clear PG feature was observed above $T_c$. However the data do show the same break in slope at $p\approx0.13$ consistent with the crossover from PG to SC gap, and retaining the coloration assists the eye to see this. These authors also noted that their gaps followed the doping dependence of the ARPES ($\pi$,0) gaps. But we now know the ARPES ($\pi$,0) gaps in the underdoped region to originate from the PG. In fact the conductance peaks show exactly the behavior expected from the square of the DOS shown in Fig.~\ref{DOS}. The coherence peaks are large for the overdoped samples where the sole gap is the SC gap, $\Delta_0$. But for the underdoped samples they have shrunk markedly (see Fig.1 of ref(\cite{Miyakawa2}). The break occurs precisely at the crossover from red to green data points in Fig.~\ref{SHgaps}(e). All four green data points have weak coherence peaks, as expected for a coexisting PG. Moreover, for the dispersion $E_{\textbf{k}}=\sqrt{\epsilon_{\textbf{k}}^2+\Delta(\textbf{k})^2+E_g(\textbf{k})^2}$, even if the pseudogap is the dominant gap, the composite gap feature will move to lower energy as $T \rightarrow T_c$ because $\Delta_0(T) \rightarrow 0$. Such a partial contraction of the gap energy near $T_c$ is not an indication the gap is a purely SC gap.

Thus we feel that the general features seen in SIN and SIS tunneling are rather consistent with all the other spectroscopic results noted above. Moreover, the picture we present is entirely consistent with recent STM tunneling data\cite{Pushp}.  What is puzzling is why the PG is not clearly seen above $T_c$ in SIN and SIS. This could be a combination of the \textbf{k}-dependent tunneling matrix elements and the fact that the ($\pi$,0) quasiparticles, which are coherent in the SC state, become incoherent in the normal state. Below $T_c$, we have previously\cite{TallonLoram} drawn attention to additional second-gap features in the tunneling spectra that could be the pseudogap ($E_g < \Delta_0$) in the case of the $p=0.16$ sample\cite{Miyakawa1}, and the SC gap ($\Delta_0 < E_g$ in the case of the $p=0.112$ sample\cite{Miyakawa2}. These are reminiscent of the second-gap features seen in Fig.~\ref{DOS} and are shown by the open green and red symbols, respectively, in Fig.~\ref{SHgaps}(e).

Our last comparison is with ARPES measurements of energy gaps. There is a very large body of ARPES literature reporting energy gaps but this is complicated by the fact that many of these are mid-point, leading-edge gaps which tend to underestimate the gap and many of the remainder are peak-to-$E_F$ energy differences which potentially overestimate the gap. The most reliable approach is to divide out the Fermi function, symmetrize the spectra and determine the
peak-to-peak gap. Even then this does not necessarily differentiate between the PG and the SC gap. This requires either measurements above and below $T_c$ or \textbf{k}-dependent measurements around the Fermi surface, or preferably both. Two papers from the Shen group\cite{Tanaka,Lee} for Bi-2212 do exactly this and thus enable clear separation of the pseudogap and the SC gap. The projected ($\pi$,0) gaps are plotted as a function of doping in Fig.~\ref{SHgaps}(f). While the scatter is quite large the same trend is seen as in all the other spectroscopies we have considered, with the PG overtaking the magnitude of the SC gap at about $p\approx0.13$. We also plot the data of Campuzano {\it et al.}\cite{Campuzano} which is the simple low-$T$ quasiparticle peak position. It therefore represents the combination of both gaps when they are both present. It also shows the same break or kink in its doping evolution.

Finally, we compare the $\Delta_0$ values with the mean-field $T_c$ value, $T_c^{mf}$, determined from the fluctuation specific heat. We have shown\cite{Fluc} that both amplitude and phase fluctuations set in simultaneously above the observed $T_c$ thus shifting $T_c$ well below $T_c^{mf}$. Using an entropy-balance procedure we found that even in the overdoped region this downward shift exceeds 10K. But in the underdoped region the shift rises to as much as 60 or 70K in the case of Bi-2212 (see $T_c$ data in Fig.~\ref{SHgaps}(d)). Significantly, while 2$\Delta/k_BT_c$ is variable and grows to large values at lower doping, 2$\Delta/k_BT_c^{mf}$ remains constant and comparable to the $d$-wave weak-coupling BCS value, 4.3. It is $T_c^{mf}$ which is the true SC energy scale, not $T_c$. In Fig.~\ref{SHgaps}(d) we plot values of $2.3\times T_c^{mf}$ and they track the $\Delta_0$ values until $E_g$ becomes comparable to $\Delta_0$ when the pseudogap is large enough to significantly reduce the order parameter and $T_c^{mf}$ then falls. Indeed, as noted, the $B_{2g}$ gap is an approximate measure of ($2\times \frac{3}{4}\times$) the SC order parameter, $\Delta_0^{\prime}$ (not, as stated, to be confused with $\Delta_0$ unless $E_g=0$). As shown by the open black circles in Fig.~\ref{RamanPG}(e) the quantity 4.3$\times$$T_c^{mf}$ follows the $B_{2g}$ gap rather closely, implying that 2$\Delta_0^{\prime}/k_BT_c^{mf}$ remains close to the weak-coupling BCS value 4.3 across the entire doping range. This is a central result. A similar comparison with the specific heat or infrared $\Delta_0$ values in Fig.~\ref{SHgaps}(a) or (b) also returns values of 2$\Delta_0/k_BT_c^{mf}$ very close to 4.3 when $E_g$ is small or absent.

The presence of SC fluctuations will cause a partial tunneling gap above $T_c$\cite{Larkin} which must not be confused with the PG. It too will have washed out coherence peaks but will only persist to $T_c^{mf}$ which is usually much less than the PG temperature, $T^*$\cite{Fluc}. It is SC fluctuations that give the impression that the SC gap evolves smoothly into the PG\cite{Harris,Fischer} and it is the SC fluctuation gap extending only 10K or so above $T_c$ that has been observed in overdoped STM data\cite{Fischer}. In both cases this has led to confusion between the SC gap and the PG. It is the fact that $\Delta_0$ and $E_g$, and $T_c$, $T_c^{mf}$ and $T^*$ are all comparable in magnitude\cite{Fluc}, combined with the presence of fluctuations above $T_c$ that has sustained this long-standing confusion which hopefully can now be laid to rest.

In conclusion, we have examined specific heat, $c$-axis infrared spectroscopy, Raman $B_{1g}$ and $B_{2g}$ scattering, intrinsic tunneling, SIS/SIN \& STM tunneling and ARPES in HTS cuprates. We find that they all present a consistent picture of two monotonically increasing energy gaps as doping is reduced. The pseudogap, $E_g$, rises rapidly from zero at $p_{crit}=0.19$ and reaches a scale of $\sim J$ at $p=0$, where $J$ is the exchange energy. It forms first around ($\pi$,0) leaving ungapped Fermi arcs on which the SC gap opens. While the Fermi arcs decrease with decreasing temperature they still remain finite in length at $T=0$. The arcs also shorten with decreasing doping, pinching off at the $d$-wave nodes at $p\approx0.05$ where SC disappears. The $T=0$ SC gap opens at $p\approx0.27$ and rises less rapidly but always monotonically, reaching a scale of about $\frac{1}{3}J$. The $B_{2g}$ Raman gap, which naively shows an apparent scaling with the dome-shaped $T_c$ curve must be reinterpreted in a two-gap scenario and the downturn is shown to be due to redistributed spectral weight. When corrected, both $B_{1g}$ and $B_{2g}$ Raman scattering present a consistent picture of a monotonically increasing SC gap. Unfortunately, H\"{u}fner {\it et al}.\cite{Hufner} reiterate this naive picture of a dome-shaped $B_{2g}$ SC gap in their review, paying no attention to the long-standing thermodynamic, NMR and (more recent) infrared data. Their resultant gap plot is further reproduced in detail by Kohsaka {\it et al}.\cite{Kohsaka} and has quickly become widely accepted. We hope the present work refutes this picture. Most importantly, $\Delta_0$ does not scale with $T_c$ but with its mean-field value, $T_c^{mf}$ across the overdoped region, with 2$\Delta_0/k_BT_c^{mf}$ remaining close to the BCS weak-coupling value of 4.3. With the opening of the pseudogap the order parameter $\Delta_0^{\prime}$ falls below $\Delta_0$ but it seems that, quite generally, 2$\Delta_0^{\prime}/k_BT_c^{mf} \approx 4.3$ at all doing levels. The cuprates may be more conventional than we once thought.


\begin{references}
\small

\bibitem{TallonLoram} Tallon J.L. \& Loram J.W. The doping dependence of $T^*$: what is the real high-$T_c$ phase
diagram?  {\it Physica C} {\bf 349}, 53-68 (2001).

\bibitem{Kanigel} Kanigel A. {\it et al.} From Fermi Arcs to the Nodal Metal: Scaling of the
Pseudogap with Doping and Temperature. {\it Nature Phys.} {\bf 2}, 447-451 (2006).

\bibitem{Miyakawa1} Miyakawa N., Zasadzinski J.F., Ozyuzer L., Guptasarma P., Hinks D.G., Kendziora C. \& Gray K.E. Predominantly Superconducting Origin of Large Energy Gaps in Underdoped Bi$_2$Sr$_2$CaCu$_2$O$_{8+d}$ from Tunneling Spectroscopy. {\it Phys. Rev. Lett.} {\bf 83}, 1018-1021 (1999).

\bibitem{Miyakawa2} Miyakawa N., Guptasarma P., Zasadzinski J.F., Hinks D.G. \& Gray K.E. Strong Dependence of the Superconducting Gap on Oxygen Doping from Tunneling Measurements on Bi$_2$Sr$_2$CaCu$_2$O$_{8-d}$. {\it Phys. Rev. Lett.} {\bf 80}, 157-160 (1998).

\bibitem{Sacuto} Le Tacon M., Sacuto A., Georges A., Kotliar G., Gallais Y., Colson D.
\& Forget A. Two energy scales and two distinct quasiparticle dynamics in the superconducting state of underdoped cuprates. {\it Nature Phys.} {\bf 2}, 537-543 (2006).

\bibitem{Guyard} Guyard W., Le Tacon M., Cazayous M., Sacuto A., Georges A.,
Colson D. \& Forget A. Breakpoint in the evolution of the gap through the cuprate phase diagram. {\it Phys. Rev. B} {\bf 77}, 024524 (2008).

\bibitem{Loram} Loram J.W., Luo J., Cooper J.R., Liang W.Y. \& Tallon J.L. Evidence on the pseudogap and condensate from the electronic specific heat. {\it J. Phys. Chem. Solids} {\bf 62}, 59-64 (2001).

\bibitem{StoreyFArc} Storey J.G., Tallon J.L. \& Williams G.V.M. Pseudogap ground state in high-temperature superconductors. {\it Phys. Rev. B} {\bf 78}, 140506(R) (2008).

\bibitem{Fluc} Tallon J.L., Storey J.G. \& Loram J.W. Fluctuations and $T_c$ reduction in cuprate superconductors. {\it Phys. Rev. Lett.} (submitted).

\bibitem{Zheng} Zheng G.Q., Kuhns P.L., Reyes A.P., Liang B. \& Lin C.T. Critical Point and the Nature of the Pseudogap of Single-Layered Copper-Oxide Bi$_2$Sr$_{2-x}$La$_x$CuO$_{6+d}$ Superconductors. {\it Phys. Rev. Lett}. {\bf 94}, 047006 (2005).

\bibitem{Mali} Mali M., Roos J., Keller H., Dooglav A.V., Sakhratov Y.A. \& Savinkov A.V. Clues obtained from the oxygen isotope effect on NMR/NQR parameters observed in YBa$_2$Cu$_4$O$_8$. {\it J. Supercon.} {\bf 15}, 511-515 (2002).

\bibitem{Pushp} Pushp A. {\it et al.} Extending universal nodal excitations optimizes superconductivity in Bi$_2$Sr$_2$CaCu$_2$O$_{8+\delta}$. {\it Science} {\bf 324}, 1689 (2009).

\bibitem{Bernhard} Bernhard C., Tallon J.L., Blasius Th., Golnik A. \& Niedermayer Ch., Anomalous Peak in the Superconducting Condensate Density of Cuprate High-$T_c$ Superconductors at a Unique Doping State. {\it Phys. Rev. Lett.} {\bf 86}, 1614-1617 (2001).

\bibitem{Yu} Yu L., Munzar D., Boris A.V., Yordanov P., Chaloupka J., Wolf Th., Lin C.T., Keimer B. \& Bernhard C. Evidence for Two Separate Energy Gaps in Underdoped High-Temperature Cuprate. {\it Phys. Rev. Lett.} {\bf 100}, 177004 (2008).

\bibitem{Pimenov} Pimenov A.V., Boris A.V., Yu Li, Hinkov V., Wolf Th., Tallon J.L., Keimer B. \& Bernhard C. Nickel Impurity-Induced Enhancement of the Pseudogap of Cuprate High-$T_c$ Superconductors. {\it Phys. Rev. Lett.} {\bf 94}, 277003 (2005).

\bibitem{Sugai} Sugai S., Suzuki H., Takayanagi Y., Hosokawa T. \& Hayamizu N. Carrier-density-dependent momentum shift of the coherent peak and the LO phonon mode in $p$-type high-$T_c$ superconductors. {\it Phys. Rev. B} {\bf 68}, 184504 (2003).

\bibitem{Hufner} H\"{u}fner S., Hossain M.A., Damascelli A. \& Sawatzky G.A. Two gaps make a high-temperature superconductor? {\it Rep. Prog. Phys.} {\bf 71}, 062501 (2008).

\bibitem{StoreyRaman} Storey J.G., Tallon J.L., Williams G.V.M. \& Loram J.W. Fermi arcs in cuprate superconductors: Tracking the pseudogap below $T_c$ and above $T^*$. {\it Phys. Rev. B} {\bf 76}, 060502(R) (2007).

\bibitem{RAMAN} Wenger F. \& Kall M. Screened Raman response in two-dimensional $d_{x^2-y^2}$-wave superconductors: Relative intensities in different symmetry channels. {\it Phys. Rev. B} {\bf 55}, 97-100 (1997).

\bibitem{Plate} Plat\'{e} M. {\it et al.} Fermi Surface and Quasiparticle Excitations of Overdoped Tl$_2$Ba$_2$CuO$_{6+d}$. {\it Phys. Rev. Lett.} {\bf 95}, 077001 (2005).

\bibitem{Limonov} Limonov M., Lee S., Tajima S. \& Yamanaka A. Superconductivity-induced resonant Raman scattering in multilayer high-$T_c$ superconductors. {\it Phys. Rev. B} {\bf 66}, 054509 (2002).

\bibitem{Krasnov} Krasnov V.M. Interlayer tunneling spectroscopy of Bi$_2$Sr$_2$CaCu$_2$O$_{8+d}$: A look from inside on the doping phase diagram of high-$T_c$ superconductors. {\it Phys. Rev. B} {\bf 65}, 140504(R) (2002).

\bibitem{Yurgens} Yurgens A., Winkler D., Claeson T., Ono S. \& Ando Y. Intrinsic Tunneling Spectra of Bi$_2$(Sr$_{2-x}$La$_x$)CuO$_{6+d}$. {\it Phys. Rev. Lett.} {\bf 90}, 147005 (2003).

\bibitem{Zavaritsky} Zavaritsky V.N. Intrinsic Tunneling or Joule Heating? {\it Phys. Rev. Lett.} {\bf 92}, 259701 (2004).

\bibitem{Krasnov2} Krasnov V.M., Sandberg M. \& Zogaj I. In situ Measurement of Self-Heating in Intrinsic Tunneling Spectroscopy. {\it Phys. Rev. Lett.} {\bf 94}, 077003 (2005).

\bibitem{Krasnov3} Krasnov V.M. Comment on "Essence of intrinsic tunneling: Distinguishing intrinsic features from artifacts". {\it Phys. Rev. B} {\bf 75}, 146501 (2007).

\bibitem{Benseman} Benseman T. \& Cooper J.R. (private communication).

\bibitem{Tanaka} Tanaka K., Lee W.S., Lu D.H., Fujimori A., Fujii T., Risdiana, Terasaki I., Scalapino D.J., Devereaux T.P., Hussain Z. \& Shen Z.-X. Distinct Fermi-Momentum-Dependent Energy Gaps in Deeply Underdoped Bi2212. {\it Science} {\bf 314}, 1910-1913 (2006).

\bibitem{Lee} Lee W.S., Vishik I.M., Tanaka K., Lu D.H., Sasagawa T., Nagaosa N., Devereaux T.P., Hussain Z. \& Shen Z.X. Abrupt onset of a second energy gap at the superconducting transition of underdoped Bi2212. {\it Nature (London)} {\bf 450}, 81-84 (2007).

\bibitem{Campuzano} Campuzano J.C. {\it et al.} Electronic spectra and their relation to the ($\pi$,$\pi$) collective mode in high-$T_c$ superconductors. {\it Phys. Rev. Lett.} {\bf 83}, 3709-3712 (1999).

\bibitem{Larkin} Larkin A. \& Varlamov A. {\it Theory of Fluctuations in Superconductors} (OUP, Oxford, 2005), p.264.

\bibitem{Harris} Harris J.M., Shen Z.-X., White P.J., Marshall D.S., Schabel M.C., Eckstein J.N. \& Bozovic I. Anomalous superconducting state gap size versus $T_c$ behavior in underdoped Bi$_2$Sr$_2$Ca$_{1-x}$Dy$_x$Cu$_2$O$_{8+d}$. {\it Phys. Rev. B} {\bf 54}, R15665-R15668 (1996).

\bibitem{Fischer} Fischer ${\O}$., Kugler M., Maggio-Aprile I., Berthod C. \& Renner C. Scanning tunneling spectroscopy of high-temperature superconductors. {\it Rev. Mod. Phys.} {\bf 79}, 353-419 (2007).

\bibitem{Kohsaka} Kohsaka Y. {\it et al}. How Cooper pairs vanish approaching the Mott insulator in Bi$_2$Sr$_2$CaCu$_2$O$_{8+d}$. {\it Nature (London)} {\bf 454}, 1072-1078 (2008).


\end{references}
\end{document}